\documentclass[9pt, conference]{IEEEtran}
\IEEEoverridecommandlockouts

\newcommand{\bfb}{{\boldsymbol b}}

\newcommand{\bfd}{{\boldsymbol d}}

\newcommand{\bfg}{{\boldsymbol g}}
\newcommand{\bfh}{{\boldsymbol h}}

\newcommand{\diag}{\mathrm{diag}}

\newcommand{\bfn}{{\boldsymbol n}}

\newcommand{\bfp}{{\boldsymbol p}}

\newcommand{\bfr}{{\boldsymbol r}}
\newcommand{\bfs}{{\boldsymbol s}}

\newcommand{\bfv}{{\boldsymbol v}}

\newcommand{\bfx}{{\boldsymbol x}}

\newcommand{\bfA}{{\boldsymbol A}} 
\newcommand{\bfB}{{\boldsymbol B}}
\newcommand{\bfC}{{\boldsymbol C}}
\newcommand{\bfD}{{\boldsymbol D}}

\newcommand{\bfM}{{\boldsymbol M}}

\newcommand{\bfP}{{\boldsymbol P}}

\newcommand{\bfR}{{\boldsymbol R}}

\newcommand{\bfT}{{\boldsymbol T}}

\newcommand{\bfW}{{\boldsymbol W}}
\newcommand{\bfX}{{\boldsymbol X}}

\newcommand{\rirs}{\boldsymbol{R}_{\text{RIS}}} 
\newcommand{\rtx}{\boldsymbol{R}_{\text{Tx}}}

\newcommand{\Phimat}{{\boldsymbol{\it{\Phi}}}} 
\newcommand{\phivec}{\pmb{\phi}} 

\newcommand{\channelT}{{\sqrt{\delta}}\rirs^{1/2} {\bfW} \rtx^{1/2, \mathrm{H}}} 

\newcommand{\dm}{\mathrm{d}}
\newcommand{\Tm}{\mathrm{T}}

\newcommand{\Hm}{{\mathrm{H}}}

\newcommand{\Rtx}{\bfR_{\text{Tx}}}
\newcommand{\Ex}{\mathop{{}\mathbb{E}}}

\newcommand{\Tmean}{\overline{{\bfT}}}

\def\BibTeX{{\rm B\kern-.05em{\sc i\kern-.025em b}\kern-.08em T\kern-.1667em\lower.7ex\hbox{E}\kern-.125emX}}
\usepackage{cite}
\usepackage{amsmath,amssymb,amsfonts}
\usepackage{graphicx}
\usepackage{textcomp}
\usepackage{xcolor}
\def\BibTeX{{\rm B\kern-.05em{\sc i\kern-.025em b}\kern-.08em
    T\kern-.1667em\lower.7ex\hbox{E}\kern-.125emX}}

\usepackage{afterpage}
\usepackage{tikz}
\usepackage{lipsum}
\usepackage{fancyhdr}

\usetikzlibrary{%
	arrows,%
	automata,%
	backgrounds,%
	calc,%
	patterns,%
	plotmarks,%
	positioning,%
	shapes.geometric,%
	shapes,%
	snakes,%
	decorations.pathmorphing,%
	decorations.shapes,%
	graphs,%
	chains,%
	arrows.meta, %
}

\usepackage{cite}
\usepackage{amsmath,amssymb,amsfonts}
\usepackage{lipsum}
\usepackage{graphicx}
\usepackage{subcaption}
\usepackage{hyperref}
\usepackage{textcomp}
\usepackage{xcolor}
\def\BibTeX{{\rm B\kern-.05em{\sc i\kern-.025em b}\kern-.08em T\kern-.1667em\lower.7ex\hbox{E}\kern-.125emX}}

\usepackage{tikz}
\usepackage{pgfplots}
\pgfplotsset{compat=newest}
\pgfplotsset{plot coordinates/math parser=false}
\newlength\figH  
\newlength\figW  
\usetikzlibrary{quotes,arrows.meta}
\usepackage{physics}
\usepackage{amssymb}
\usepackage{amsmath}
\usepackage{amsthm}
\usepackage{float}
\usepackage{algorithm}
\usepackage{algpseudocode}
\usepackage{algorithm,caption}

\usepackage[english]{babel}

\usepackage{graphicx}
\usepackage{wrapfig}
\usepackage{caption}
\usepackage{array}
\usepackage{multirow}
\usepackage{arydshln}
\usepackage{color}
\usepackage{xcolor}
\usepackage{tcolorbox}
\usepackage[framemethod=tikz]{mdframed}
\fancypagestyle{IEEEtitlepagestyle}{
    \fancyhf{} 
    \fancyfoot[L]{\vspace{12pt}%
    \footnotesize
    \copyright This work has been submitted to the IEEE for possible publication. Copyright may be transferred without notice, after which this version may no longer be accessible.}
}
\begin{document}
\title{Low Complexity Rate Splitting Approach in RIS-Aided Systems Based on Channel Statistics\\
}

\author{\IEEEauthorblockN{Sadaf Syed, Michael Joham, Wolfgang Utschick }
\IEEEauthorblockA{{School of Computation, Information and Technology, Technical University of Munich, Germany}\\
Emails: \{sadaf.syed, joham, utschick\}@tum.de}
}



\maketitle

\begin{abstract}
Rate splitting multiple access~(RSMA) and reconfigurable intelligent surface~(RIS) are two prospective technologies for improving the spectral and energy efficiency in future wireless communication systems. In this work, we investigate a rate splitting~(RS) technique for an RIS-aided system in the presence of only statistical channel knowledge. We propose an algorithm with a quasi closed-form solution based only on the second-order channel statistics, which reduces the design complexity of the system as it does not require estimation of the channel state information~(CSI) and optimisation of the precoding filters and phase shifts of the RIS in every channel coherence interval.
\end{abstract}

\begin{IEEEkeywords}
Downlink, RIS, Rate Splitting, Statistical CSI, MISO
\end{IEEEkeywords}

\section{Introduction}
Rate splitting multiple access~(RSMA) and reconfigurable intelligent surface~(RIS) are two key technologies that have received considerable attention for B5G/6G systems. Rate splitting~(RS) mitigates the interference effectively by splitting the messages of the users into common and private parts \cite{rs1, rs2, rs3, rs4, rs5, amor2024efficient, amor2023rate}. The common parts are jointly encoded into a common stream, which is decoded by all users, and the private parts are encoded into the respective private streams, which are decoded only by the corresponding user. Therefore, RS allows the users to partially decode the interference and partially treat it as noise. On the other hand, RIS is another prospective low-cost technology that can enhance the spectral efficiency of B5G/6G systems \cite{wu2019intelligent, guo2020weighted, zhang2020capacity, semmler2022linear, syed2023design, syed2024design, hu2020statistical, dang2020joint, zhang2022sum, noh2022cell}. Specifically, an RIS is a passive array composed of a large number of reflecting elements, where the phase shifts can be configured to achieve the desired system's performance. Owing to the performance benefits of both RS and RIS, many recent works have investigated the integration of the two technologies \cite{rsris1, rsris2, rsris3, fang2022fully, rsris4 ,ge2023rate}.
\par The existing algorithms on the combination of RS and RIS mostly rely on the assumption of perfect channel state information~(CSI) \cite{rsris1, rsris2, rsris3, fang2022fully, rsris4}. The method in \cite{ge2023rate} considers channel statistics but it also assumes perfect CSI for the channel between the BS and the RIS. Owing to the passive structure of the RIS, the acquisition of perfect CSI for the RIS-associated links is very time-consuming~\cite{joham2022estimation}. Moreover, the algorithms in \cite{rsris1, rsris2, rsris3, fang2022fully, rsris4} require the beamforming vectors and the phase shifts of the RIS to be optimised in every channel coherence interval, which is computationally very expensive because both RS and RIS require demanding optimisation algorithms. In this work, we propose an algorithm which is only based on the channel statistics. Since the structure of the channels varies slowly, the covariance matrices remain constant for many channel coherence intervals. Hence, it is possible to obtain accurate information of the second-order statistics of the channels through long-term observation. Note that the systems which are designed based on the covariance matrices need to be optimised only once per coherence interval of the covariance matrices, which is much larger than the channel coherence interval (cf. \cite{syed2023design, syed2024design, hu2020statistical, dang2020joint, zhang2022sum, noh2022cell}). Moreover, these algorithms do not require channel estimation in every channel coherence interval, which significantly reduces the channel training overhead and the design complexity. The proposed algorithm, which integrates both RS and RIS, maximises the achievable sum-rate of the users using only the channel statistics and without any knowledge of the actual CSI. Moreover, it offers a quasi closed-form solution which alleviates the need of using interior point solvers like CVX \cite{grant2014cvx}, as used in \cite{rsris1, rsris2,fang2022fully, rsris4}. Additionally, we also present an alternative approach that considers the imperfect channel knowledge in the optimisation process, which is performed in every channel coherence interval. This algorithm provides a benchmark for comparing our algorithm based only on statistical CSI. 

\section{System Model}
The downlink~(DL) of an RIS-aided multi-user multiple-input single-output (MISO) time division duplex~(TDD) system is investigated. The system is equipped with one base station~(BS) having $M$ antennas and serving $K$ single-antenna users. The RIS consists of $N$ passive reflecting elements. The phase-shift matrix is defined by the diagonal matrix $\Phimat$ = $\diag$($\phi_1,\cdots,\phi_N$), where $\phi_1, \cdots, \phi_N$ are the phase shifts of the $N$ elements of the RIS, and $\phivec = \diag(\Phimat)$ denotes the phase-shift vector. The channel from the BS to the $k$-th user is denoted by $\bfh_{\dm, k}\in \mathbb{C}^{M}$ which is complex Gaussian distributed with zero mean, such that $\bfh_{\dm,k}\sim\mathcal{N_\mathbb{C}}({\bf{0}}, \bfC_{\dm,k})$. The channel from the RIS to the $k$-th user is denoted by $\bfr_{k}\in \mathbb{C}^{N}$ with $\bfr_{k}\sim\mathcal{N_\mathbb{C}}({\bf{0}}, \bfC_{\bfr, k})$. The channel from the BS to the RIS is assumed to follow the Kronecker channel model with non-zero mean~\cite{kermoal2002stochastic}, given by
\begin{align}
{\bfT} = \left(\sqrt{1 - \delta}\right)\bfT' + \channelT \label{eqn1}
\end{align}
where the mean $\left(\sqrt{1 - \delta}\right)\bfT' = \Tmean$ denotes the deterministic line-of-sight~(LoS) component between the BS and the RIS. The entries in $\bfW\in \mathbb{C}^{N \times M}$ are independently and identically distributed~(i.i.d.) with unit variance and zero mean following complex Gaussian distributions. $\bfR_{\text{RIS}}$ and $\bfR_{\text{Tx}}$ denote the channel correlation matrices on the side of the RIS and the BS respectively, and $\delta \in [0,1]$ is the factor controlling the strength of the deterministic LoS component.    
The effective channel for the $k$-th user is given by
\begin{align}
\label{eqn4.1}
\bfh_{k}^{\Hm} & = \bfh_{\dm,k}^{\Hm} + \bfr_{k}^{\Hm} \Phimat^{\Hm}\bfT
\end{align}
where $\bfh_{\dm,k}, \:\bfT$ and $\bfr_{k}$ are mutually independent. It is clear from \eqref{eqn4.1} that $\bfh_k$ is non-Gaussian (as it contains the product of Gaussian random variables) with zero-mean and its covariance matrix is denoted by $\bfC_k$, which is given by (see \cite[ Sec. III]{syed2024design}) 
\begin{align}
   &\bfC_k  = \Ex\left[\bfh_k \bfh_k^{\Hm} \right] = \bfC_{\dm,k} + \Tmean^{\Hm}\Phimat\bfC_{\bfr,k}\Phimat^{\Hm}\Tmean + \delta \phivec^{\mathrm{H}} \bfX_k \phivec \rtx  \label{eqnC}  \\ 
    & \text{where}\:\bfX_k = \rirs \odot \bfC_{\bfr, k}^{\Tm}, \text{and $\odot$ denotes the Hadamard product.} \nonumber
 \end{align}
In the data transmission phase, we assume that a single layer RS approach is applied at the BS such that the message intended for a user is split into one common message and one
private message. 
\section{Problem Formulation and Solution}
At the BS, the message for the $k$-th user is split into a common part and a private part. The common parts of all the users are combined and encoded into a common message $s_\text{c}$. The private parts are independently encoded into the private messages $s_1, \cdots, s_K$. We denote $\bfs = [s_\text{c}, s_1, \cdots , s_K]^\Tm \in \mathbb{C}^{K+1}$ and $\bfP = [\bfp_\text{c}, \bfp_1,\cdots , \bfp_K] \in \mathbb{C}^{M \times (K+1)}$ as the data stream vector and the beamforming matrix respectively, where $\bfp_\text{c}$ is the common precoding vector and $\bfp_1, \cdots, \bfp_K$ are the private beamforming vectors.
We assume that each stream in $\bfs$ is zero-mean with unit variance, i.e., $\Ex[{\bfs \bfs^{\Hm}}]=\mathbf{I}$. The total received signal at the $k$-th user is given by
\begin{align}
    y_{k} =\bfh_{k}^{\Hm}\bfp_{\text{c}}\:s_{\text{c}} + \sum\limits_{i = i }^K \bfh_{k}^{\Hm}\bfp_{i}\:s_{i} + z_{k}
\end{align}
where $z_{k} \sim\mathcal{N_\mathbb{C}}(0, 1) $ denotes the additive white Gaussian noise (AWGN) at the $k$-th user's side. Our objective is to reduce the training complexity, and hence, to design the system based on the channel statistics only because the coherence interval of the covariance matrices is much longer than the channel coherence interval. The ergodic sum rate of the RSMA system can be expressed as
\begin{align}
    R_{\text{erg}} = \Ex \left[\sum\limits_{i=1}^{K}\log_2\left(1 + \gamma_{p,i}  \right) \right] + \!\min\limits_{k} \:\Ex \Big[\log_2\left(1 + \gamma_{\text{c},k}  \right)\Big] \label{ex}
\end{align}
\begin{align}
\text{with}\quad \gamma_{p,i}  = \dfrac{\big|{\bfh_{i}^{\Hm}}{\bfp_{i}}\big|^2}{\sum\nolimits_{\substack{j \neq i}}\big|{\bfh_{i}^{\Hm}}{\bfp_{j}}\big|^2 + 1}, \quad \gamma_{\text{c},k}  = \dfrac{\big|{\bfh_{k}^{\Hm}}{\bfp_{\text{c}}}\big|^2}{\sum\nolimits_{\substack{j  }}\big|{\bfh_{k}^{\Hm}}{\bfp_{j}}\big|^2 + 1} 
\end{align}
which are the signal-to-interference-noise ratio~(SINR) of the private and common parts respectively. The inner minimisation over the user indices $k$ in \eqref{ex} ensures that the common message is decodable by all users. 
Using Lemma 1 of \cite{approxbound}, the expectation terms in \eqref{ex} can be approximated as 
\begin{align}
   R_{\text{erg}} \approx  \bar{R}_{\text{erg}} =  \sum\limits_{i=1}^{K}\log_2\left(1 + \bar{\gamma}_{p,i}  \right) + \!\min\limits_{k} \log_2\left(1 + \bar{\gamma}_{\text{c},k}  \right)\label{approx}
\end{align}
\begin{align}
\label{lbp}
\bar{\gamma}_{p,i} &= \dfrac{\mathop{{}\mathbb{E}}\left[\left|\bfh_i^{\Hm}\bfp_i\right|^2\right]}{\sum\nolimits_{\substack{j \neq i}}{\mathop{{}\mathbb{E}}\left[\left|{{\bfh_i^{\Hm}}{\bfp_j}}\right|^2\right] + 1}} = \dfrac{\bfp_i^{\Hm}\bfC_i \bfp_i}{\sum\nolimits_{\substack{j \neq i}}{ \bfp_j^{\Hm}\bfC_i \bfp_j + 1}}
\end{align}
\begin{align}
\label{lbc}
\bar{\gamma}_{\text{c},k} &= \dfrac{\mathop{{}\mathbb{E}}\left[\left|\bfh_k^{\Hm}\bfp_\text{c}\right|^2\right]}{\sum\nolimits_{\substack{j }}{\mathop{{}\mathbb{E}}\left[\left|{{\bfh_k^{\Hm}}{\bfp_j}}\right|^2\right] + 1}} = \dfrac{\bfp_\text{c}^{\Hm}\bfC_k \bfp_\text{c}}{\sum\nolimits_{\substack{j }}{ \bfp_j^{\Hm}\bfC_k \bfp_j + 1}}
\end{align}
where $\bar{\gamma}_{p,i}$ and $\bar{\gamma}_{\text{c},k}$ are the approximations of the SINRs. The approximation in Lemma 1 of \cite{approxbound} has been used in many works, e.g., \cite{gan2021ris, zhi2022power, fan2015uplink}. The optimisation problem can hence be formulated as 
\begin{align}
&\!\max\limits_{\bfP, \phivec}  &  & \sum\limits_{i=1}^{K}\log_2\left(1 + \bar{\gamma}_{p,i}  \right) \: +  \:\!\min\limits_{k} \log_2\left(1 + \bar{\gamma}_{\text{c},k}  \right)\tag{P1} \label{P1}\\
& \quad \text{s.t.}  &     &   \sum\limits_{i=1}^{K}\|{\bfp_i}\|^2 + \|{\bfp_\text{c}}\|^2\leq P_t, \: |\phi_n| = 1 \:\forall \:n = 1, \cdots, N  \nonumber. 
\end{align}
The first constraint is due to the DL power constraint at the BS and the second constraint is the unit-modulus constraint of the phase shifts of RIS. Since the optimisation problem \eqref{P1} depends only on the channel statistics, $\bfP$ and $\phivec$ designed by \eqref{P1} also only depend on the channel statistics, and hence, they are kept fixed in the coherence interval of the covariance matrices.    
\subsection{Simplification of the Optimisation Problem}
It is difficult to solve \eqref{P1} in closed-form, especially due to the search of the minimum common rate among all users. To simplify \eqref{P1}, we use the max-min inequality \cite[Ch. 5]{boyd2004convex} and hence, the objective function in \eqref{P1} can be expressed as 
\begin{align}
\!\max\limits_{\bfP, \phivec}   \quad \!\min\limits_{k} \quad \sum\limits_{i=1}^{K} \log_2\left(1 + \bar{\gamma}_{p,i}  \right) +   \log_2\left(1 + \bar{\gamma}_{\text{c},k}  \right) \nonumber
\end{align}
\begin{align}
 \leq \!\min\limits_{k}   \quad \!\max\limits_{\bfP, \phivec}  \quad \sum\limits_{i=1}^{K} \log_2\left(1 + \bar{\gamma}_{p,i}  \right) +   \log_2\left(1 + \bar{\gamma}_{\text{c},k}  \right). \label{reform}
\end{align}
Hence, \eqref{P1} can be suboptimally solved by optimising its upper bound in \eqref{reform}. To this end, the inner maximisation problem is firstly solved for each user index $k$ and then, the best user index is selected by solving the outer minimisation problem. The objective function in \eqref{reform} is still very involved, which can be further simplified by the fractional programming~(FP) approach of \cite{fractional1, fractional2} to a more tractable form. Following the similar steps as in \cite[Sec. III]{syed2024design}, the objective function can be rewritten as
{\small
\begin{align}
    &\log\left(1 + \lambda_{\text{c},k} \right) - \lambda_{\text{c},k}  + 2\left(\sqrt{1 + \lambda_{\text{c},k}}\right)\text{Re}\left\{{\boldsymbol{\beta}}_{\text{c},k}^{\Hm}\bfC_{k}^{1/2, \Hm}\bfp_{\text{c}}\right\} \nonumber \\
    &- \|{\boldsymbol{\beta}}_{\text{c},k}\|_2^2 \Bigg(\sum\limits_{j=1}^{K}\bfp_j^{\Hm}\bfC_{k} \bfp_j + \bfp_\text{c}^{\Hm}\bfC_{k} \bfp_\text{c} + 1 \Bigg) \nonumber \\
    &+ \sum\limits_{i=1}^{K}\left(\log\left(1 + \lambda_i \right) - \lambda_i + 2\left(\sqrt{1 + \lambda_i}\right)\text{Re}\left\{{\boldsymbol{\beta}}_i^{\Hm}\bfC_i^{1/2, \Hm}\bfp_i\right\} \right) \nonumber \\& - \sum\limits_{i=1}^{K}\|{\boldsymbol{\beta}}_i\|_2^{2} \Bigg(\sum\limits_{j=1}^{K}\bfp_j^{\Hm}\bfC_i \bfp_j + 1 \Bigg)  \label{fp}
\end{align}
}
where $\lambda_i$, $\lambda_{\text{c},i}$ , ${\boldsymbol{\beta}}_i \in \mathbb{C}^{M}$ and ${\boldsymbol{\beta}}_{\text{c},i} \in \mathbb{C}^{M}$ are the auxiliary variables of the FP approach. The inner maximisation problem of \eqref{reform} is solved for the $k$-th user by maximising the reformulated function in \eqref{fp} using the iterative non-convex BCD method \cite{bcd, bcdmain}, where $\bfP$, $\phivec$ and the auxiliary variables are alternatingly updated in each iteration. 
\subsection{Solution Approach using only Statistical CSI }
The closed-form solution of the auxiliary variables alternatingly maximising the objective function in \eqref{fp} reads as \cite[Sec. IV]{syed2024design}
\begin{align}
    \lambda_i = \bar{\gamma}_{p,i},   \quad \lambda_{\text{c},k} = \bar{\gamma}_{\text{c},k} \label{lamk}
\end{align}
\begin{align}
   {\boldsymbol{\beta}}_i =  \dfrac{\left(\sqrt{1 + \lambda_i}\right)\bfC_i^{1/2, \Hm} \bfp_i}{\sum\nolimits_{j = 1 }^{K}{ \bfp_j^{\Hm}\bfC_i \bfp_j + 1}}   \label{betak1}
   \end{align}
   \begin{align}
   {\boldsymbol{\beta}}_{\text{c},k} =  \dfrac{\left(\sqrt{1 + \lambda_{\text{c},k}}\right)\bfC_k^{1/2, \Hm} \bfp_\text{c}}{\sum\nolimits_{j = 1 }^{K}{ \bfp_j^{\Hm}\bfC_k \bfp_j + \bfp_\text{c}^{\Hm}\bfC_k \bfp_\text{c} + 1}}.  \label{betak}
\end{align}
Next, for updating the beamforming matrix $\bfP$, we first denote $\bfv = \text{vec}(\bfP)$ and introduce selection matrices $\bfD_\text{c}, \bfD_1,\cdots, \bfD_K \in \mathbb{C}^{M \times (K+1)M}$ such that  $\bfp_\text{c} = \bfD_\text{c} \bfv$ and $\bfp_i = \bfD_i \bfv$. The selection matrices are given by
\begin{align*}
    \bfD_\text{c} = \left[{\mathbf{I}}_M,  {\mathbf{0}}_M, \cdots, {\mathbf{0}}_M\right], \:\: \bfD_i = [ \underbrace{{\mathbf{0}}_M,\cdots}_{i.M}, {\mathbf{I}}_M,  \cdots, {\mathbf{0}}_M].  
\end{align*}
Optimisation of $\bfP$ also requires satisfying the DL power constraint at the BS, and it can be efficiently handled by applying the rescaling trick mentioned in \cite{joham2002transmit, christensen2008weighted, zhao2023rethinking}. The idea is to introduce the constraint into the objective function itself by exploiting the fact that the inequality constraint in \eqref{P1} will be satisfied with equality for the optimal precoding filters (see \cite[Sec. IV]{syed2024design}). With $\|\bfv\|_2^{2} = P_t$, the simplified objective function for the update of precoders is given by 
\begin{align}
   \!\min\limits_{k}   \quad \!\max\limits_{\bfv}  \quad -\bfv^{\Hm}\bfA_k \bfv + 2 \text{Re}\{ \bfb_k^{\Hm} \bfv \} + u_k \label{objf}
\end{align}
{\small
\begin{align}
   &\text{where} \quad \bfA_k =  \sum\limits_{i=1}^{K}\|{\boldsymbol{\beta}}_i\|_2^{2} \Bigg(\sum\limits_{j=1}^{K}\bfD_j^{\Hm}\bfC_i \bfD_j + \dfrac{1}{P_t}{\mathbf{I}} \Bigg)  \nonumber \\ & \quad+ \|{\boldsymbol{\beta}}_{\text{c},k}\|_2^{2} \Bigg(\sum\limits_{j=1}^{K}\bfD_j^{\Hm}\bfC_k \bfD_j + \bfD_\text{c}^{\Hm}\bfC_k \bfD_\text{c} +\dfrac{1}{P_t}{\mathbf{I}} \Bigg), 
   \end{align}
   }
   {\small
\begin{align}
   & \quad \bfb_k^{\Hm} = \sum\limits_{i=1}^{K} \left(\sqrt{1 + \lambda_i}\right){\boldsymbol{\beta}}_i^{\Hm}\bfC_i^{1/2, \Hm}\bfD_i \nonumber +  \left(\sqrt{1 + \lambda_{\text{c},k}}\right){\boldsymbol{\beta}}_{\text{c},k}^{\Hm}\bfC_{k}^{1/2, \Hm}\bfD_\text{c} \nonumber 
   \end{align}
   \begin{align}
     \text{and}\quad u_k =  \log\left(1 + \lambda_{\text{c},k} \right) - \lambda_{\text{c},k}.
\end{align}
}
Note that $\bfA_k$ is a positive definite matrix which makes the inner maximisation block in \eqref{objf} a concave function of $\bfv$, and hence, its global optimal solution depending on $k$ can be obtained as
\begin{align}
    \bfv_k' = \bfA_k^{-1} \bfb_k.  \label{v}
\end{align}
Even though the DL power constraint has been incorporated into the objective function of \eqref{objf}, the optimal solution in \eqref{v} must be rescaled to ensure that the constraint is satisfied with equality (cf.~\cite{joham2002transmit, christensen2008weighted, zhao2023rethinking}). Therefore, the rescaled solution is given by 
\begin{align}
    \bfv_k = \dfrac{\sqrt{P_t}}{\| \bfA_k^{-1} \bfb_k \|_2}\bfA_k^{-1} \bfb_k.  \label{v2}
\end{align}
Finally, after obtaining the optimal $\bfv_k$ for each $k$ by \eqref{v2}, we can obtain the optimal user index $k_{\text{opt}}$ which minimises $-\bfv_k^{\Hm}\bfA_k \bfv_k + 2 \text{Re}\{ \bfb_k^{\Hm} \bfv_k \} + u_k$, thereby yielding $\bfv_{\text{opt}} = \bfv_{k_{\text{opt}}}$. 
\par For the update of $\phivec$, we need to reformulate \eqref{fp} in terms of $\phivec$ using the expression of $\bfC_k$ from \eqref{eqnC}. Denoting $\bar{\phivec} = [\phivec, 1]^{\Tm}$ and after some reformulation steps, we get the following optimisation problem
\begin{align}
   \!\min\limits_{k}   \quad \!\max\limits_{\bar{\phivec}}  \quad \bar{\phivec}^{\Hm} \left(\bfB + \bfB_{\text{c},k} \right) \bar{\phivec}  + t_k, \quad \text{s.t.}\:|\bar{\phi}_n| = 1 \:\forall \:n \label{phiop}
\end{align}
where the expressions of $\bfB$, $\bfB_{\text{c},k}$ and $t_k$ are provided in the Appendix (see~\eqref{mat}). Ignoring the constraints in \eqref{phiop}, the optimal $\bar{\phivec}_k$ for the inner maximisation problem for each user index $k$ is given by the principal eigenvector of the matrix $\bfB+\bfB_{\text{c},k}$. The obtained $\bar{\phivec}_k$ is then normalised to satisfy the unit-modulus constraint, which makes the solution suboptimal but it still leads to a good performance as shown in the simulation results later. Finally, the optimal user index $k_{\text{opt}}$ is chosen which minimises the objective function in \eqref{phiop}, and we get $\bar{\phivec}_{\text{opt}} = \bar{\phivec}_{{k_{\text{opt}}}}$. \eqref{P1} is thus solved suboptimally by replacing its objective function by \eqref{fp} and alternatingly applying \eqref{lamk}, \eqref{betak1}, \eqref{betak}, \eqref{v2}, and heuristically solving \eqref{phiop} till the convergence is reached.
\subsection{Complexity of the Algorithm}
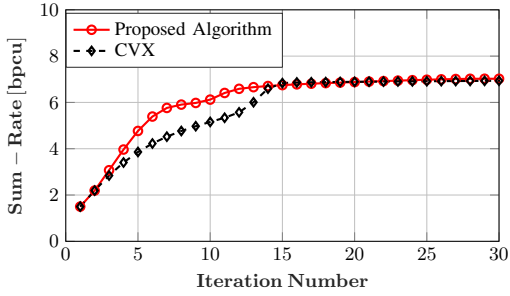
\begin{figure}		
		\scalebox{0.8}{
%
%
\definecolor{mycolor1}{rgb}{0.00000,0.44700,0.74100}%
\definecolor{mycolor2}{rgb}{0.85000,0.32500,0.09800}%
\definecolor{mycolor3}{rgb}{0.92900,0.69400,0.12500}%
\definecolor{mycolor4}{rgb}{0.49400,0.18400,0.55600}%
\definecolor{mycolor5}{rgb}{0.46600,0.67400,0.18800}%
\definecolor{mycolor6}{rgb}{0.30100,0.74500,0.93300}%
\definecolor{mycolor7}{rgb}{0.63500,0.07800,0.18400}%
\begin{tikzpicture}

\begin{axis}[%
width=0.9\figW,
height=\figH,
at={(0\figW,0\figH)},
scale only axis,
xmin=0,
xmax=30,
xlabel style={font=\color{white!15!black}},
xlabel={$\bf{Iteration\:Number}$},
ymin=0,
ymax=10,
ylabel style={font=\color{white!15!black}},
ylabel={$\bf{Sum-Rate\:[bpcu]}$},
axis background/.style={fill=white},
xmajorgrids,
ymajorgrids,
legend style={at={(0,0.75)}, anchor=south west, legend cell align=left, align=left, draw=white!15!black, row sep=-0.05cm},
legend columns=1
]
\addplot [color=red, solid, line width=1.0pt, mark=o, mark options={solid, red}]
  table[row sep=crcr]{%
1	1.49901612289934\\
2	2.1939322575676\\
3	3.07144648177202\\
4	3.96464275185284\\
5	4.76736042344374\\
6	5.38975335407578\\
7	5.76165509583673\\
8	5.90585729666963\\
9	5.97704459666155\\
10	6.12137056854559\\
11	6.40823737986834\\
12	6.58734298456461\\
13	6.65890578082207\\
14	6.70816242327668\\
15	6.74743236260836\\
16	6.7802754471859\\
17	6.80852107534932\\
18	6.83379185176188\\
19	6.85673054151038\\
20	6.87860379928647\\
21	6.89980213620995\\
22	6.92067917195718\\
23	6.94125808756642\\
24	6.96120695561707\\
25	6.9799183469639\\
26	6.99689997295758\\
27	7.01071523624008\\
28	7.02007846117579\\
29	7.02258254275823\\
30	7.02193260125975\\
31	7.02620735262261\\
32	7.03144015235801\\
33	7.03711056158259\\
34	7.04316823055202\\
35	7.04922960169724\\
36	7.04922960169724\\
};
\addlegendentry{Proposed Algorithm}

\addplot [color=black, dashed, line width=1.0pt, mark=diamond, mark options={solid, black}]
  table[row sep=crcr]{%
1	1.49901612289934\\
2	2.19200395121372\\
3	2.84071839818803\\
4	3.40413522213871\\
5	3.85938473288834\\
6	4.2248945429485\\
7	4.52116942387929\\
8	4.76507216237765\\
9	4.9725254271406\\
10	5.15729355878787\\
11	5.3372962433562\\
12	5.57205042349753\\
13	6.00753052129735\\
14	6.59622336461952\\
15	6.83396254789871\\
16	6.86336317226106\\
17	6.87084317420926\\
18	6.87717552964583\\
19	6.8832793757662\\
20	6.88913678592294\\
21	6.89473063462511\\
22	6.90006012901017\\
23	6.90513772521755\\
24	6.90997641277047\\
25	6.91459112826109\\
26	6.91899590509996\\
27	6.92320460072181\\
28	6.92723040807846\\
29	6.93108581665575\\
30	6.93478325583957\\
31	6.93833354323561\\
32	6.94174600291\\
33	6.94503052170972\\
34	6.94819606904922\\
35	6.95125093476202\\
36	6.95420275737819\\
37	6.95705863178787\\
38	6.95982508299337\\
39	6.96250798351737\\
40	6.96511302425635\\
41	6.96764538266228\\
42	6.97010970240944\\
43	6.97251034921144\\
44	6.97485119070429\\
45	6.97713556646736\\
46	6.97936682084053\\
47	6.98154743655703\\
48	6.98368096712481\\
49	6.98576972247942\\
50	6.98781592576981\\
51	6.98982161107425\\
52	6.99178862569542\\
53	6.99371865951199\\
54	6.99561326359316\\
55	6.99747385720257\\
56	6.99930173271034\\
57	7.00109807298869\\
58	7.00286400924619\\
59	7.00459969426493\\
60	7.0063071509319\\
61	7.00798686306647\\
62	7.0096396645747\\
63	7.01126629603592\\
64	7.01286717070923\\
65	7.01444374852349\\
66	7.01599555053785\\
67	7.01752329015356\\
68	7.01902749542167\\
69	7.02050865793686\\
70	7.02196722862334\\
71	7.02340364440862\\
72	7.02481984913273\\
73	7.02621463849916\\
74	7.02758838567895\\
75	7.02894145035018\\
76	7.03027417997268\\
77	7.03158691119271\\
78	7.03287996725918\\
79	7.03415366272754\\
80	7.03540829078972\\
81	7.03664409517008\\
82	7.03786142508921\\
83	7.03906055733366\\
84	7.04024179211113\\
85	7.04140543398063\\
86	7.04255178202564\\
87	7.04368114221288\\
88	7.04479385594531\\
89	7.04589037872978\\
90	7.04697071372617\\
91	7.04803518192574\\
92	7.04908414600607\\
93	7.05011789304984\\
94	7.05113676610958\\
95	7.05214099990032\\
96	7.05313075549667\\
97	7.05410635104418\\
98	7.05506805587696\\
99	7.05601613721702\\
100	7.056950864826\\
101	7.05787247457371\\
102	7.05878111972851\\
103	7.05967717434491\\
104	7.06056087485126\\
105	7.0614326102503\\
106	7.06229236667668\\
107	7.06314052833982\\
108	7.06397727963487\\
109	7.06480272054496\\
110	7.06561728097839\\
111	7.06642109485401\\
112	7.06721444174637\\
113	7.06799705443414\\
114	7.06876982729028\\
115	7.06953269635564\\
116	7.0702858058774\\
117	7.07102933791696\\
118	7.07176346731736\\
119	7.07248836283114\\
120	7.07320380148181\\
121	7.07391057152226\\
122	7.07460859752091\\
123	7.07529811998805\\
124	7.07597925627791\\
125	7.07665216497294\\
126	7.07731699678747\\
127	7.07797389864488\\
128	7.07862301374927\\
129	7.07926513945695\\
130	7.07989944752368\\
131	7.08052632813129\\
132	7.08114593184993\\
133	7.08175839810056\\
134	7.08236385926452\\
135	7.08296244267565\\
136	7.08355427149915\\
137	7.08413946534048\\
138	7.08471814163681\\
139	7.08529041397823\\
140	7.08585639520007\\
141	7.08641619458023\\
142	7.08696991840756\\
143	7.08751767117787\\
144	7.08805955600314\\
145	7.08859568834836\\
146	7.08912616347493\\
147	7.089651075871\\
148	7.0901705188971\\
149	7.09068458417587\\
150	7.09119336243997\\
151	7.09169694260273\\
152	7.09219541144343\\
153	7.09268885381345\\
154	7.09317735315371\\
155	7.0936609923117\\
156	7.09413985364069\\
157	7.09461401750494\\
158	7.09508355971718\\
159	7.09554855544096\\
160	7.09600908027802\\
161	7.09646520800051\\
162	7.09691701212219\\
163	7.09736456398807\\
164	7.09780793350061\\
165	7.09824719042163\\
166	7.09868240244323\\
167	7.09911363591364\\
168	7.09954095617758\\
169	7.099964427189\\
170	7.10038411171072\\
171	7.10080007202834\\
172	7.1012123684944\\
173	7.10162106383613\\
174	7.10202621124456\\
175	7.10242786949555\\
176	7.10282609661376\\
177	7.1032209467821\\
178	7.10361247483004\\
179	7.10400073423266\\
180	7.10438577732898\\
181	7.10476765588398\\
182	7.10514644514123\\
183	7.10552214916875\\
184	7.10589481416164\\
185	7.1062644949046\\
186	7.10663124636987\\
187	7.1069951226696\\
188	7.10735617575272\\
189	7.10771445683435\\
190	7.1080700090534\\
191	7.10842287517517\\
192	7.10877309945819\\
193	7.10912072689018\\
194	7.10946579833379\\
195	7.10980835607814\\
196	7.11014844032566\\
197	7.11048608993537\\
198	7.11082134231856\\
199	7.11115423333934\\
200	7.11148497947975\\
};
\addlegendentry{CVX}

\end{axis}
\end{tikzpicture}%
} 
		\caption{Convergence Plot for $K=3$, $M = 4$, $N = 40$}
		\label{conv}
\end{figure}

The complexity of the proposed algorithm is dominated by computing $K$ times the inverse of matrices $\bfA_k \in \mathbb{C}^{(K+1)M \times (K+1)M }$ in every iteration. The principal eigenvectors of matrices~$\bfB + \bfB_{\text{c},k} \in~\mathbb{C}^{(N+1) \times (N+1)}$ can be found efficiently with the power iteration method \cite[Ch. 27]{trefethen2022numerical}. However, note that the entire process needs to be performed just once in every coherence interval of the covariance matrices, i.e., we do not need to update $\bfP$ or $\phivec$ whenever the channels change. Hence, the proposed algorithm is much cheaper compared to those in  \cite{rsris1, rsris2, rsris3, fang2022fully, rsris4}, which require optimisation in every channel coherence interval. Moreover, since the algorithm has a quasi closed-form solution for each step (we observed less than 20 steps till convergence, as in Fig.~\ref{conv}), it is computationally cheap. Problem $\eqref{P1}$ (after its reformulation to \eqref{fp}) could also be solved by the interior-point solvers like CVX \cite{grant2014cvx} directly without using the max-min inequality of \eqref{reform} for finding the minimum common rate. But the run-time using CVX is around 100 times more than our algorithm. The convergence plot in Fig. \ref{conv} for $P_t$ = 10 dB illustrates that the proposed algorithm converges to a similar value as the CVX approach despite being computationally much less expensive. 
\section{Algorithm using Imperfect CSI }
In this section, we introduce another approach for comparing our proposed algorithm. This approach is based on a lower bound of the achievable sum-rate which considers the imperfect channel knowledge \cite{hassibi2003much}. The application of the bound to the RS approach without an RIS has already been investigated in \cite{amor2024efficient}. The optimisation problem using this bound is given by 
{
\begin{align}
&\!\max\limits_{\bfP, \phivec}  &  & \sum\limits_{i=1}^{K}\log_2\left(1 + \hat{\gamma}_{p,i}  \right) \: +  \:\!\min\limits_{k} \log_2\left(1 + \hat{\gamma}_{\text{c},k}  \right)  \label{P2} \tag{P2} 
\end{align}
\begin{align}
& \quad \text{s.t.}  &     &   \sum\limits_{i=1}^{K}\|{\bfp_i}\|^2 + \|{\bfp_\text{c}}\|^2\leq P_t, \: |\phi_n| = 1 \:\forall \:n = 1, \cdots, N \nonumber
\end{align}
\begin{align}
\label{lbp2}
\text{with }\hat{\gamma}_{p,k}  = \dfrac{| \hat{\bfh}_k^{\Hm}\bfp_k|^2}{\sum\nolimits_{j = 1 }^{K}{ \bfp_j^{\Hm}\bfC_{\text{err},k} \bfp_j + \sum\nolimits_{j \neq k}| \hat{\bfh}_k^{\Hm}\bfp_j|^2+ 1}}
\end{align}
\begin{align}
\label{lbc2}
\hat{\gamma}_{\text{c},k}  = \dfrac{| \hat{\bfh}_k^{\Hm}\bfp_\text{c}|^2}{\bfp_\text{c}^{\Hm}\bfC_{\text{err},k} \bfp_\text{c}  + \sum\nolimits_{j = 1 }^{K}{ \bfp_j^{\Hm}\bfC_{\text{err},k} \bfp_j + \sum\nolimits_{j = 1 }^{K}| \hat{\bfh}_k^{\Hm}\bfp_j|^2+ 1}}
\end{align}
}
where $\hat{\bfh}_k$ is the least squares~(LS) channel estimate given by $\hat{\bfh}_k = \bfh_k + \bfn_k$, and $\bfn_k$ is the zero-mean channel estimation error whose covariance matrix is denoted by $\bfC_{\text{err},k} = \Ex[\bfn_k \bfn_k^{\Hm}]$. Note that the lower bound in \cite{hassibi2003much} requires the channel estimate to be the minimum mean square error~(MMSE) estimate. However, due to the non-Gaussianity of $\bfh_k$, it is very difficult to compute the MMSE estimate for systems with RIS \cite{joham2022estimation} and hence, we pragmatically employ LS estimate $\hat{\bfh}_k$. The bound in \eqref{lbp2} and \eqref{lbc2} with the LS estimate is no more a lower bound but it is just an approximation to the actual sum-rate. \eqref{P2} can be reformulated using the FP approach in the same way as the proposed algorithm in Section~III, where \eqref{lbp} and \eqref{lbc} are replaced by \eqref{lbp2} and \eqref{lbc2} respectively. The reformulated problem has a form similar to \eqref{fp}, which can then be solved either by CVX or the approach described in Section III. The details of the derivation have been omitted here due to the lack of space. This approach using the imperfect channel knowledge requires the update of the beamforming matrix and the phase shifts of the RIS in every channel coherence interval, thereby making it computationally more expensive than our algorithm based only on statistical CSI. However, it leads to a better performance in the presence of imperfect CSI than naively using the algorithms designed for the perfect CSI case.  
\section{Results and Discussion}
In this section, numerical results are provided to validate the effectiveness of the proposed algorithm. The channels are generated according to their distributions described in Section~II. The covariance matrix of each channel is generated according to the 3GPP specifications \cite{etsi5138}. The details of the channel modelling and generation of covariance matrices are given in \cite{syed2024design}. The proposed algorithm in Section III using only the channel statistics is denoted by {\it{Stat CSI}} whereas the algorithm considering the imperfect channel knowledge in Section IV is denoted by {\it{Imp CSI}}. We compare the algorithms with different settings (with/without RS and with/without RIS). The algorithms with optimised phase shifts are denoted by {\it{OptRIS}}, those with random phase shifts by {\it{RandRIS}}, and the ones without RIS by {\it{noRIS}}. The cases without RS ({\it{noRS}}) are obtained by switching off the common rate part of the algorithms. The sum-rate of the users is plotted in Figs.~\ref{rate1} and~\ref{rate2} over different transmit power levels $P_t$ for $K = 3, M = 4$, $N = 40$ and $K=10, M=8$, $N = 60$ respectively. The sum-rate is averaged over 1000 channel realisations for 100 different covariance matrices, which are generated by varying the position of the users and the path powers. As expected, using both RS and optimised phase shifts gives the best performance. It is also observed that the CVX approach yields similar results as the proposed algorithm. We compare the performance of the proposed algorithm with the method in \cite{syed2024design}, which uses a generalised matched filter~(GMF) with imperfect CSI. The GMF-based approach in \cite{syed2024design} uses RIS without RS, and its sum-rate starts saturating after 10~dB (also observed in \cite{syed2024design}). We also compare the algorithms with the RS approach in \cite{amor2024efficient} without RIS, which considers the MMSE channel estimates and is based on the lower bound of \cite{hassibi2003much}, similar to {\it{Imp~CSI}}. For Fig.~\ref{rate1}, the channels are estimated with an SNR of $0$ dB, which is constant even over different $P_t$ since the users' transmit power is fixed in the TDD systems. It is observed in Fig.~\ref{rate1} that the {\it{Stat CSI}} RS algorithm without RIS performs better than \cite{amor2024efficient} at higher $P_t$ even though \cite{amor2024efficient} is computationally more expensive and requires optimisation in every channel coherence interval. This can be attributed to the fact that the channel estimation error is high compared to $P_t$ at high transmit power levels. The same algorithm in Fig.~\ref{rate2} performs much better when the channels are estimated at a higher SNR level of $25$ dB. We next compare the proposed {\it{Stat CSI}} algorithm with the approaches using perfect CSI~\cite{fang2022fully} (single RIS case) and imperfect CSI (section IV) in Fig.~\ref{imp}, where the estimation error covariance matrix of the effective channel is $\bfC_{\text{err, k}} = {\mathbf{I}} \quad \forall k$. Note that the total variance of the estimation error is equally divided among $\bfh_{d,k}$ and the $N$ cascaded channel vectors for each element of RIS, i.e., each channel vector is estimated at an SNR of 16 dB approximately. As expected, the algorithm in \cite{fang2022fully}, based on the perfect CSI performs the best when the channels are known without any error. However, if the algorithm is naively used in the presence of channel estimation errors without any modification, it suffers from a significant performance deterioration. {\it{Imp CSI}} of Section IV, which considers the channel estimation error, performs well in the presence of imperfect CSI, and it performs better than the proposed {\it{Stat CSI}} algorithm at high $P_t$ but at the cost of much higher complexity because it needs optimisation of both precoders and RIS in every channel coherence interval. It can also be seen from Fig.~\ref{imp} that the {\it{Imp CSI}} approach without RS starts saturating at high $P_t$, illustrating the significance of RS in the scenarios of imperfect CSI.
\begin{figure}		
		\scalebox{0.8}{
%
%
\definecolor{mycolor1}{rgb}{1.00000,0.00000,1.00000}%
\definecolor{mycolor3}{rgb}{0.92900,0.69400,0.12500}%
\definecolor{mycolor4}{rgb}{0.49400,0.18400,0.55600}%
\definecolor{mycolor5}{rgb}{0.46600,0.67400,0.18800}%
\definecolor{mycolor6}{rgb}{0.30100,0.74500,0.93300}%
\definecolor{mycolor7}{rgb}{0.63500,0.07800,0.18400}%
\begin{tikzpicture}

\begin{axis}[%
width=0.9\figW,
height=\figH,
at={(0\figW,0\figH)},
scale only axis,
xmin=-10,
xmax=40,
xlabel style={font=\color{white!15!black}},
xlabel={$\bf{Transmit\:Power\:{\it{P_t}}\:in\:dB}$},
ymin=0,
ymax=20,
ylabel style={font=\color{white!15!black}},
ylabel={$\bf{Sum-Rate\:[bpcu]}$},
axis background/.style={fill=white},
xmajorgrids,
ymajorgrids,
legend style={at={(0,0.99)}, anchor=south west, legend cell align=left, align=left, draw=white!15!black, row sep=-0.05cm, font=\scriptsize},
legend columns=2
]
\addplot [color=black, solid, line width=1.0pt, mark=o, mark options={solid, black}]
  table[row sep=crcr]  {%
-10	3.13790531539709\\
10	9.15779395126351\\
20	12.3522561085383\\
30	16.0500307437084\\
40	19.8012682778709\\
};
\addlegendentry{Stat CSI RS + OptRIS}

\addplot [color=green, dashdotted, line width=1.0pt, mark=diamond, mark options={dashdotted, green}]
  table[row sep=crcr]  {%
-10	3.220263501970687 \\
10	9.076350197689369\\
20	12.526184918304610\\
30	15.681305800952314\\
40	19.071722454871697\\
};
\addlegendentry{Stat CSI RS + OptRIS (CVX)}

\addplot [color=red, dashed, line width=1.0pt, mark=diamond, mark options={solid, red}]
  table[row sep=crcr]{%
-10	0.0127756562797343\\
10	0.770397799345244\\
20	2.2699894891751\\
30	5.27161467309845\\
40	8.88497360370418\\
};
\addlegendentry{Stat CSI RS + noRIS}

\addplot [color=blue, line width=1.0pt, mark=star, mark options={solid, blue}]
  table[row sep=crcr]{%
-10	1.70883218849647\\
10	5.68442033747678\\
20	8.31700132662526\\
30	11.9495626235936\\
40	15.4845134505288\\
};
\addlegendentry{Stat CSI RS + RandRIS}

\addplot [color=red, line width=1.0pt, mark=o, mark options={solid, red}]
  table[row sep=crcr]{%
-10	0.00542553326974078\\
10	0.534218110065813\\
20	1.626480321495\\
30	4.42925922702764\\
40	7.09391222683526\\
};
\addlegendentry{Stat CSI noRS + noRIS}

\addplot [color=orange, dashed, line width=1.0pt, mark=o, mark options={solid, orange}]
  table[row sep=crcr]{%
-10	0.875086341079697\\
10	2.89276307622501\\
20	5.90185013441439\\
30	9.17884117154812\\
40	12.3055863746518\\
};
\addlegendentry{Stat CSI noRS + RandRIS}

\addplot [color=mycolor6]
  table[row sep=crcr]{%
-10	1.39652263345668\\
10	4.09291510555332\\
20	7.33228717897073\\
30	10.0135589762891\\
40	13.9108663429897\\
};
\addlegendentry{Stat CSI noRS + OptRIS}

\addplot [color=mycolor7]
  table[row sep=crcr]{%
-10	1.82143596420979\\
10	3.96374405147531\\
20	4.64690969083956\\
30	4.74103575235236\\
40	4.74764925218098\\
};
\addlegendentry{Stat CSI GMF \cite{syed2024design}}


\addplot [color=cyan, dashdotted, line width=1.0pt, mark=diamond, mark options={solid, cyan}]
  table[row sep=crcr]{%
-10	0.0116688312861507\\
10	0.72000891354574\\
20	2.27197562419727\\
30	4.55648248740297\\
40	6.8111307876327\\
};
\addlegendentry{Imp CSI RS + noRIS \cite{amor2024efficient}}

\end{axis}
\end{tikzpicture}%

} 
		\caption{Sum Rate vs $P_t$ in dB for $K=3$, $M = 4$, $N = 40$}
		\label{rate1}
\end{figure}
\begin{figure}		
		\scalebox{0.8}{
%
%
\definecolor{mycolor1}{rgb}{1.00000,0.00000,1.00000}%
\definecolor{mycolor2}{rgb}{0.85000,0.32500,0.09800}%
\definecolor{mycolor3}{rgb}{0.92900,0.69400,0.12500}%
\definecolor{mycolor4}{rgb}{0.49400,0.18400,0.55600}%
\definecolor{mycolor5}{rgb}{0.46600,0.67400,0.18800}%
\definecolor{mycolor6}{rgb}{0.30100,0.74500,0.93300}%
\definecolor{mycolor7}{rgb}{0.63500,0.07800,0.18400}%
\begin{tikzpicture}

\begin{axis}[%
width=0.9\figW,
height=\figH,
at={(0\figW,0\figH)},
scale only axis,
xmin=-10,
xmax=40,
xlabel style={font=\color{white!15!black}},
xlabel={$\bf{Transmit\:Power\:{\it{P_t}}\:in\:dB}$},
ymin=0,
ymax=20,
ylabel style={font=\color{white!15!black}},
ylabel={$\bf{Sum-Rate\:[bpcu]}$},
axis background/.style={fill=white},
xmajorgrids,
ymajorgrids,
legend style={at={(0,0.90)}, anchor=south west, legend cell align=left, align=left, draw=white!15!black, row sep=-0.05cm, font=\scriptsize},
legend columns=2
]
\addplot [color=black, solid, line width=1.0pt, mark=o, mark options={solid, black}]
  table[row sep=crcr]{%
-10	2.22635364159498\\
10	7.84618251555712\\
20	11.4754209725097\\
30	14.8608927685729\\
40	17.8262404344159\\
};
\addlegendentry{Stat CSI RS + OptRIS}

\addplot [color=red, dashed, line width=1.0pt, mark=diamond, mark options={solid, red}]
  table[row sep=crcr]{%
-10	0.0202402944781628\\
10	1.03395425799507\\
20	3.01838931276457\\
30	6.1220669932156 \\ 
40	10.03123235392581 \\ 
};
\addlegendentry{Stat CSI RS + noRIS}

\addplot [color=blue, line width=1.0pt, mark=star, mark options={solid, blue}]
  table[row sep=crcr]{%
-10	2.06116167539173\\
10	4.20184947215208\\
20	6.6275999274849\\
30	9.68598847913377\\
40	12.4048742458174\\
};
\addlegendentry{Stat CSI RS + RandRIS}

\addplot [color=red, line width=1.0pt, mark=o, mark options={solid, red}]
  table[row sep=crcr]{%
-10	0.00506797083879235\\
10	0.496416005032698\\
20	2.12305452954947\\
30	5.8220669932156\\
40	9.23370366704425\\
};
\addlegendentry{Stat CSI noRS + noRIS}

\addplot [color=orange, dashed, line width=1.0pt, mark=o, mark options={solid, orange}]
  table[row sep=crcr]{%
-10	0.933771209577552\\
10	2.00745530211037\\
20	3.83626250395572\\
30	6.95018651052885\\
40	10.1224262824743\\
};
\addlegendentry{Stat CSI noRS + RandRIS}

\addplot [color=mycolor6]
  table[row sep=crcr]{%
-10	1.28341400273223\\
10	3.06491296445922\\
20	5.71128596530024\\
30	9.2919883329908\\
40	12.5048085109366\\
};
\addlegendentry{Stat CSI noRS + OptRIS}

\addplot [color=mycolor7]
  table[row sep=crcr]{%
-10	2.06208208939699\\
10	4.69888013180371\\
20	5.62336017970394\\
30	6.08725432900185\\
40	6.12982546412445\\
};
\addlegendentry{Stat CSI GMF \cite{syed2024design}}

\addplot [color=mycolor1, dashdotted, line width=1.0pt, mark=star, mark options={solid, mycolor1}]
  table[row sep=crcr]{%
-10	0.0270008008698219\\
10	1.35573818638952\\
20	4.10260446332053\\
30	8.52562010968859\\
40	11.6342698202185\\
};
\addlegendentry{Imp CSI RS + noRIS \cite{amor2024efficient}}

\end{axis}

\end{tikzpicture}%
}
		\caption{Sum Rate vs $P_t$ in dB for $K=10$, $M = 8$, $N = 60$}
		\label{rate2}
\end{figure}
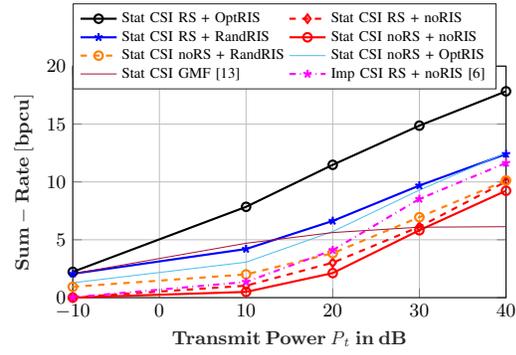 
\begin{figure}		
		\scalebox{0.8}{
%
%
\definecolor{mycolor1}{rgb}{1.00000,0.00000,1.00000}%
\definecolor{mycolor2}{rgb}{0.85000,0.32500,0.09800}%
\definecolor{mycolor3}{rgb}{0.92900,0.69400,0.12500}%
\definecolor{mycolor4}{rgb}{0.49400,0.18400,0.55600}%
\definecolor{mycolor5}{rgb}{0.46600,0.67400,0.18800}%
\definecolor{mycolor6}{rgb}{0.30100,0.74500,0.93300}%
\definecolor{mycolor7}{rgb}{0.63500,0.07800,0.18400}%
\begin{tikzpicture}

\begin{axis}[%
width=0.9\figW,
height=\figH,
at={(0\figW,0\figH)},
scale only axis,
xmin=-10,
xmax=30,
xlabel style={font=\color{white!15!black}},
xlabel={$\bf{Transmit\:Power\:{\it{P_t}}\:in\:dB}$},
ymin=0,
ymax=30,
ylabel style={font=\color{white!15!black}},
ylabel={$\bf{Sum-Rate\:[bpcu]}$},
axis background/.style={fill=white},
xmajorgrids,
ymajorgrids,
legend style={at={(0,0.97)}, anchor=south west, legend cell align=left, align=left, draw=white!15!black, row sep=-0.05cm, font=\scriptsize},
legend columns=2
]

\addplot [color=red, line width=1.0pt, mark=o, mark options={solid, red}]
  table[row sep=crcr]{%
-10	6.38199096655857\\
0	9.8649949141171\\
10	15.143919570283\\
20	21.5974550220377\\
30	28.0772128348051\\
};
\addlegendentry{\cite{fang2022fully} with perfect CSI RS + OptRIS }

\addplot [color=black, solid, line width=1.0pt, mark=o, mark options={solid, black}]
  table[row sep=crcr]{%
-10	6.16299742465678\\
0	8.88802013329962\\
10	13.2620255434678\\
20	18.762993935221\\ 
30	24.7522986119488\\
};
\addlegendentry{Imp CSI RS + OptRIS}

\addplot [color=red, dashed, line width=1.0pt, mark=diamond, mark options={solid, red}]
  table[row sep=crcr]{%
-10	4.99141097494044\\
0	8.12309645387079\\
10	11.4289701263343\\
20	13.6462180050701\\
30	17.7087421474728\\
};
\addlegendentry{Stat CSI RS + OptRIS}

\addplot [color=green, line width=1.0pt, mark=star, mark options={solid, green}]
  table[row sep=crcr]{%
-10	6.06439639841249\\
0	8.49662451080853\\
10	10.9384787747037\\
20	11.8719158808087\\
30	12.2694042511741\\
};
\addlegendentry{Imp CSI noRS + OptRIS}

\addplot [color=pink, line width=1.0pt, mark=o, mark options={solid, pink}]
  table[row sep=crcr]{%
-10	6.15745881737286\\
0	7.63918294119193\\
10	9.10748946344672\\
20	9.96322051696944\\
30	10.11806153567399\\
};
\addlegendentry{\cite{fang2022fully} with imperfect CSI RS + OptRIS}

\addplot [color=mycolor1, dashdotted, line width=1.0pt, mark=star, mark options={solid, mycolor1}]
  table[row sep=crcr]{%
-10	0.1298\\
0	0.8551\\
10	3.3236\\
20	7.6420\\
30	11.4968\\
};
\addlegendentry{Imp CSI RS + noRIS \cite{amor2024efficient}}

\end{axis}
\end{tikzpicture}%
} 
		\caption{Sum Rate vs $P_t$ in dB for $K=3$, $M = 4$, $N = 40$} 
		\label{imp}
\end{figure}
\appendix
We need to reformulate \eqref{fp} in terms of $\phivec$ using the expression of $\bfC_k$ from \eqref{eqnC}. To this end, additional variables are introduced which are given by
{\small
\begin{align*}
   &\bfx_{1,k} = \bfC_{d,k}^{1/2, \Hm} \bfp_k, \quad \bfx_{2,k} = \bfC_{r,k}^{1/2, \Hm} \Phimat^{\Hm}\Tmean \bfp_k, \: l_k =  {\delta (\bfp_k^{\Hm}\Rtx\bfp_k)},
   \end{align*}
   }
   {\small
   \begin{align*}
    &l_\text{c} ={\delta (\bfp_\text{c}^{\Hm}\Rtx\bfp_\text{c})}, \quad \bfx_{3,k} = \sqrt{l_k}\bfX_{k}^{1/2, \Hm} \phivec, \quad \bfx_{1,k,c} = \bfC_{d,k}^{1/2, \Hm} \bfp_\text{c}, 
    \end{align*} 
    \begin{align*}
    & \bfx_{2,k,c} = \bfC_{r,k}^{1/2, \Hm} \Phimat^{\Hm}\Tmean \bfp_\text{c}, \: \bfx_{3,k, c} = \sqrt{l_\text{c}}\bfX_{k}^{1/2, \Hm} \phivec, \: f = \sum\nolimits_{j}\bfp_j^{\Hm}\Rtx \bfp_j.
\end{align*}
}
Before updating $\phivec$, we decompose the numerators in \eqref{lbp} and \eqref{lbc} into 3 parts using $\bfC_k$ from \eqref{eqnC} and introduce auxiliary variables ${\boldsymbol{\beta}}_{m,k}$ and ${\boldsymbol{\beta}}_{\text{c},m,k}$ ($m \in \{1,2,3\}$) to decouple the numerators and denominators, which are alternatingly updated by the FP steps as
{\small
  \begin{align*}
   &  {\boldsymbol{\beta}}_{m,k} =  \dfrac{\left(\sqrt{1 + \lambda_k}\right)\bfx_{m,k}}{\sum\nolimits_{j }{ \bfp_j^{\Hm}\bfC_k \bfp_j + 1}}, \quad {\boldsymbol{\beta}}_{\text{c},m, k} =  \dfrac{\left(\sqrt{1 + \lambda_{\text{c},k}}\right)\bfx_{m,k,c} }{\sum\nolimits_{j }{ \bfp_j^{\Hm}\bfC_k \bfp_j + \bfp_\text{c}^{\Hm}\bfC_k\bfp_\text{c} + 1}}. 
\end{align*}
   }
   Introducing more variables for the problem reformulation, we have
   {\small
   \begin{align*}
  & a_k = \sum\nolimits_{m=1}^{3} \|{\boldsymbol{\beta}}_{m,k}\|_2^{2}, \quad a_{\text{c},k} = \sum\nolimits_{m=1}^{3} \|{\boldsymbol{\beta}}_{\text{c},m,k}\|_2^{2},  \\ &\bfM_{j,k} = \left(\Tmean \bfp_j \bfp_j^{\Hm}\Tmean^{\Hm}\right)  \odot \bfC_{r,k}^{\Tm}, \quad \bfd_k = \bfC_{r,k}^{1/2}{\boldsymbol{\beta}}_{2,k}.
   \end{align*}
   }
  After updating ${\boldsymbol{\beta}}_{m,k}$ and ${\boldsymbol{\beta}}_{\text{c},m, k}$, they are kept fixed during the update of $\phivec$. Using all the above variables, \eqref{fp} can be rewritten as a function of $\phivec$ given in \eqref{phiop}, where the matrices $\bfB$ and $\bfB_{\text{c},k}$ are 
   \begin{align} \label{mat}
   \bfB = 
\begin{bmatrix}
  -\bfR & \bfg \\
  \bfg^{\Hm} & 0
\end{bmatrix} \quad \text{and} \quad  \bfB_{\text{c},k} = 
\begin{bmatrix}
  -\bfR_{\text{c},k} & \bfg_{\text{c},k} \\
  \bfg_{\text{c},k}^{\Hm} & 0
\end{bmatrix}
 \end{align}
 and $t_k$ contains the terms in \eqref{fp} which are independent of $\phivec$. The elements of $\bfB$ and $\bfB_{\text{c},k}$ are given by
   {\small
   \begin{align*}  
   &\bfg^{\Hm} = \sum\nolimits_{j=1}^{K}\left( \sqrt{1 + \lambda_{j}}\right)\left( \bfp_j^{\Hm}\Tmean^{\Hm} \diag(\bfd_j) + \sqrt{l_j} {\boldsymbol{\beta}}_{3,j}^{\Hm} \bfX_{j}^{1/2, \Hm}\right) \\
   & \bfR = \sum\nolimits_{j=1}^{K}\left(a_j\sum\nolimits_{i=1}^{K}\bfM_{i,j} + \delta \: f \: a_j \bfX_j \right)\\
   &\bfg_{\text{c},k}^{\Hm} = ( \sqrt{1 + \lambda_{\text{c},k}})\left( \bfp_\text{c}^{\Hm}\Tmean^{\Hm} \diag(\bfd_{\text{c},k}) + \sqrt{l_\text{c}} {\boldsymbol{\beta}}_{\text{c},3,k}^{\Hm} \bfX_{k}^{1/2, \Hm}\right) \\
   & \bfR_{\text{c},k} = a_{\text{c},k}\left(\sum\nolimits_{j=1}^{K}\bfM_{j,k} + \bfM_{\text{c},k} \right)  + \delta  (f + \bfp_\text{c}^{\Hm}\Rtx\bfp_\text{c}) a_{\text{c},k} \bfX_k. 
\end{align*}
}

\bibliographystyle{IEEEtran}
\bibliography{bibliography}

\end{document}